\documentclass[a4paper, parskip=full,11pt, bibliography=totocnumbered]{scrartcl}
\input{paper.sty}

\newcommand{\inj}{\mathcal{I}}
\newcommand{\mdm}{m_{DM}}

\newcommand{\jcap}[1]{\hat{J}_{ #1 }^{DM}}
\newcommand{\Nev}{N_\text{ev}^{60}}

\newcommand{\vn}{{\mathbf{n}}}
\newcommand{\bn}{\mathbf{n}}

\begin{document}

\title{\textcolor{RoyalPurple}{Ultra High Energy Cosmic Rays \& Super-heavy Dark Matter}\vspace{.5em}}

\subtitle{Observables and prospect of (non)detection
\vspace{2em}}

\author[1,2]{Luca Marzola
\thanks{luca.marzola@ut.ee}}

\author[2]{Federico R. Urban
\thanks{federico.urban@kbfi.ee}\vspace{1em}}

\affil[1]{Laboratory of Theoretical Physics, Institute of Physics, University of Tartu; Ravila 14c, 50411 Tartu, Estonia.} 

\affil[2]{Laboratory of High Energy and Computational Physics, National Institute of Chemical Physics and Biophysics, R\"avala pst. 10, 10143 Tallinn, Estonia.}

\date{Dated: \today}

\maketitle
	\begin{center}
		\textbf{Abstract:}
	\end{center}
\begin{abstract}
\noindent We reanalyse the prospects for upcoming Ultra-High Energy Cosmic Ray experiments in connection with the phenomenology of Super-heavy Dark Matter. We identify a set of observables well suited to reveal a possible anisotropy in the High Energy Cosmic Ray flux induced by the decays of these particles, and quantify their performance via Monte Carlo simulations that mimic the outcome of near-future and next-generation experiments. 
The spherical and circular dipoles are able to tell isotropic and anisotropic fluxes apart at a confidence level as large as $4\sigma$ or $5\sigma$, depending on the Dark Matter profile. The forward-to-backward flux ratio yields a comparable result for relatively large opening angles of about 40~deg, but it is less performing once a very large number of events is considered. We also find that an actual experiment employing these observables and collecting 300~events at 60~EeV would have a $50\%$ chance of excluding isotropy against Super-heavy Dark Matter at a significance of at least $3\sigma$.
\end{abstract}

\newpage

\section{Introduction} 
\label{sec:Introduction}
Keeping up with a long-standing tradition in the literature about ``Dark Matter'' (DM from now on), we kick off by sternly pointing out the major lacuna in our understanding of the Universe that the origin of DM is. In fact, measurements of the cosmic microwave background power spectrum and the rotation curves of galaxies, amongst others, suggest the existence of a new substance which behaves like ordinary matter as far as gravity is concerned~\cite{Olive:2016xmw,Bull:2015stt}. The problem is that, unlike baryonic matter which aggregates in planets and stars, this mysterious material does not shine any light -- hence the differently creative name of DM. Since nothing in the Standard Model of Particle Physics (SM in short) has the right properties to play the r\^ole of DM, these observations are regarded as a rather solid evidence for the existence of New Physics. DM has been a central problem for Physics in the latest four decades, as testified by the over 14.000~dedicated arXiv papers. The proposed DM candidates range from condensates of scalar fields that coherently oscillate according to their potential (axion/axion-like particles)~\cite{Kim:1986ax} to sterile neutrinos or new fermions~\cite{Kim:2008pp,Boehm:2014hva,Hektor:2014kga}, including primordial black holes~\cite{Chen:2002tu,Frampton:2010sw} and particles emerging from extensions of General Relativity~\cite{Babichev:2016bxi,Babichev:2016hir,Aoki:2016zgp}.

This landscape of DM models, however, is not uniformly populated; because of the possible connections with Supersymmetric theories and the simplicity of its dynamics, one particular scenario of DM has been dominating the scene: Weakly Interacting Massive Particle (WIMP) DM~\cite{Jungman:1995df,Bertone:2004pz}. Undoubtedly, the WIMP paradigm has had a profound impact on shaping the ongoing experimental physics program dedicated to DM searches.  In spite of the extensive effort, however, so far no experiment has provided any evidence in support of WIMP DM. New data release are progressively relegating the models to unpleasant corners of their parameter spaces~\cite{Gaskins:2016cha,Akerib:2015rjg,Khachatryan:2014rra,ATLAS:2012ky} and, with no supersymmetric particle detected at the LHC~\cite{Aad:2014pda,Aad:2011hh,Chatrchyan:2011zy,Chatrchyan:2013mys}, the WIMP paradigm begins to crack and wither.

For this reason, in this paper we take a pragmatic step into the darkness of DM, focusing on a model that typically yields little signatures: Super-heavy DM (or SHDM to be original). The idea behind SHDM is that DM is composed by a gravitationally produced non-thermal relic of supermassive particles~\cite{Chung:1998zb,Kuzmin:1998uv,Kuzmin:1998kk,Kolb:1998ki,Kuzmin:1999zk}. According to the present literature, SHDM with a mass scale $\mathcal{O}(10^{12}$--$10^{14})$~GeV could originate in Supersymmetry-breaking and Kaluza-Klein models~\cite{Kolb:2007vd,Hikage:2008sk,Chung:2011xd,Chung:2011ck}, or Multi-Graviton theories~\cite{Babichev:2016bxi,Babichev:2016hir,Aoki:2016zgp}, as well as in Coleman-Weinberg extensions of the SM in concomitance with inflation~\cite{Kannike:2016jfs}. The weakness of gravitational interactions then explains the lack of any signal in direct detection experiment, whereas collider experiments would need to reach energies and luminosities significantly higher than those of current machines to probe the scheme. If SHDM particles are stable, there is no way for current and next-generation DM indirect detection experiments to prove their properties. Yet, Ultra High-Energy Cosmic Rays (UHECR) experiments may have something to say about (slowly) decaying SHDM\footnote{A complementary approach to bound the properties of SHDM candidates consists in studying their impact on inflationary observables \cite{Chung:2001cb, Aloisio:2015lva,Chung:2013rda} and on the Cosmic Microwave Background Radiation \cite{Chung:1999ve}.}.

If SHDM decays produce SM particles, it is not difficult to see that the amount of UHECR resulting from these processes should increase as we look toward SHDM denser regions, like the centre of our Galaxy. Consequently, it is possible to probe SHDM models with UHECR experiments by measuring anisotropies in the UHECR flux~\cite{Dubovsky:1998pu,Kachelriess:2003rv,Kim:2003th,Aloisio:2006yi,Aloisio:2007bh}. Motivated by this basic observation, we set out to find the best possible way to detect a SHDM signal via its features across the sky. In this paper, we thus introduce different observables aimed at revealing a possible anisotropic UHECR component and quantify their performance via Monte Carlo (MC) simulations that capture the gist of current and next generation experiments. On top of that, we quantify the power of an experiment to reject the hypothesis of an isotropic UHECR flux against a predicted SHDM one through the best performing observables.

The paper is organised as follows: the next section offers an overview of the quantities relevant for UHECR physics and specifies the ingredients from SHDM which are present in our analyses. Section~\ref{sec:Methodology} describes our observables and presents the details of the simulation algorithms. The most important results of our work are presented in Section~\ref{sec:Results}, whereas complementary ones are included in the Appendix.  We close with Section~\ref{sec:Conclusion}.

\section{An overview} 
\label{sec:An overview}
The flux of UHECR that arrive on Earth with energy
\([E, E+\td E]\) from an infinitesimal cone of aperture \(\beta\)
centred on the direction \(\vn\) can be generally decomposed into a galactic and an extra-galactic components
\begin{equation}
	J_x (E, \vn) = J_x^g (E, \vn) + J_x^{eg} (E, \vn);\qquad x\in\{p,\nu,\gamma\}\,
	\label{eq:oldeq1}.
\end{equation}
Here $x$ specifies the UHECR primary: $p$ stands for protons, $\nu$ for neutrinos, and $\gamma$ for photons. Throughout the following we take the corresponding fluxes $J_x$ to be normalised in units of area, time, energy and solid angle. For simplicity, and also because that is what is expected from SHDM decays, we will disregard all baryons but protons, however it is straightforward to extend the present analysis to heavier nuclei.

The relative contribution of protons, neutrinos and other species to the detected (or expected) flux changes significantly depending on the energy range we observe. This behaviour is due to the interactions of the UHECR injected at the source with the cosmic background and intervening media~\cite{Kampert:2014eja,Allard:2011aa,LetessierSelvon:2011dy}; for instance:
\begin{itemize}
\item
  protons with an energy roughly above 60~EeV (\textit{GZK
  limit}~\cite{Greisen:1966jv,Zatsepin:1966jv}) scatter on the Cosmic Microwave Background photons and produce \(\Delta\) resonances
  which decay into protons/neutrons and pions. The mean
  free path associated to this interaction is $\mathcal{O}(100)$~Mpc
  (\textit{GZK horizon}).
\item
  the mean free path of photons above the GZK limit is restricted to a
  few Mpc by interactions with the radio background. 
\item
  neutrinos are not affected by the cosmic 
  backgrounds but they can hardly initiate the observed air showers due to their
  reduced cross sections on nucleons.
\end{itemize}

By comparing these limits with the average size of Galaxies $\mathcal{O}(100)$~kpc, the average distance of Galaxies in a cluster $\mathcal{O}(1)$~Mpc, and the average distance of clusters in superclusters $\mathcal{O}(10)$~Mpc, we understand how tiny we are and that UHECR detected with energy above the GZK cutoff are most likely charged nuclei generated in the local supercluster. Interestingly, the palatable astrophysical sources of UHECRs contained within the GZK horizon do not typically possess the right characteristics to accelerate these particles at the observed high energies~\cite{Kotera:2011cp,LetessierSelvon:2011dy}. Another possibility is that UHECRs come from the decay of SHDM particles within our own Milky Way, which is the idea we pursue here. Thus, in the following, we will assume that the UHECR flux is dominated by the decay products of local SHDM. 

\subsection{The galactic component} 
\label{sub:The galactic component}
The galactic component of the flux is defined as

\begin{equation}
	J^g_x(E,\vn)=\frac{1}{4\pi} \int \inj_x(E, r, \vn)\, \td r;\,\qquad x\in\{p,\nu,\gamma\}
	\label{eq:galcomp}
\end{equation}

where the $4\pi$ normalisation accounts for the isotropy of the DM decay processes, and $r$ is our distance to the source located within the Milky Way. These are characterised by the intrinsic DM emissivity
\begin{equation}
	\inj_x(E, r, \vn)=n_{DM}(r, \vn) \,\frac{\td \Gamma_x}{\td E}
\end{equation}
which depends on the galactic DM density profile, $n_{DM}(r, \vn)$ and on the specific DM interactions encapsulated in the differential decay widths $\td\Gamma_x/\td E$ into the particle species $x$.

The SHDM decay into hadrons is complicated by the hadronisation process. For instance, considering a specific hadron \(h\), we have~\cite{Sarkar:2001se,Aloisio:2003xj,Basu:2004pv}
\begin{equation}
	\frac{1}{\Gamma}\frac{\td \Gamma_h}{\td x}=\sum_a\int\limits_x^1\frac{1}{\Gamma_a}\,D_a^h(z,\mu^2)\frac{\td\Gamma_a(y, \mu^2, m_{DM}^2)}{\td y}\Bigg\vert_{y=x/z}\frac{\td z}{z}
	\label{eq:decaywidths}
\end{equation}
where $m_{DM}$ is the SHDM mass that sets the energy scale of the decay process, $\Gamma:=\sum_x \Gamma_x$ and in the integral \(x\), $y$ and \(z\) are various fractions of available maximum momentum and primary parton momentum carried by the hadron under scrutiny:
\begin{equation}
	x\simeq2\frac{E_h}{m_{DM}};\qquad
	z\simeq \frac{E_h}{E_a};\qquad
	y=\frac x z\simeq2\frac{E_a}{m_{DM}}\,.
\end{equation}
Eq.~\eqref{eq:decaywidths} allows us to separate the details of hadronisations process from the fundamental properties of SHDM described in DM models. In fact, the former are contained in the \emph{fragmentation function} \(D_a^h(z, \mu)\)~\cite{Sarkar:2001se,Aloisio:2003xj,Basu:2004pv}, which essentially corresponds to the probability that a process initiated by a parton \(a\) result in a specific hadron \(h\). Notice that the while the fragmentation functions depend on the factorisation scale \(\mu\), the resulting differential branching ratio into hadron does not depend on this quantity. This forces the cancellation of the dependence on the factorisation scale order by order in perturbation theory amongst the terms on the right-hand side of the above equation, in a way that the fragmentation functions obey the so-called DGLAP equation~\cite{Dokshitzer:1977sg,Gribov:1972ri,Lipatov:1974qm,Altarelli:1977zs}. Once the fragmentation functions are measured at the electroweak scale, we can use the DGLAP equation to evolve them up to the DM scale, where the relevant decay process takes place. 

As the fragmentation functions are computed according to the above prescription, we can focus on the particle physics which regulates the remaining terms on the right-hand side of Eq.~\eqref{eq:decaywidths}. In particular, the (exclusive) differential decay width of SHDM into a parton $a$ can be computed as 
\begin{equation}
	\td \Gamma_a = \frac{\modu{\emm_a}^2}{8\pi \,m_{DM}}\delta\left(\mdm-2E_a\right)\td E_a\quad
	\implies\quad
	\frac{\td \Gamma_a}{\td E_a} = \frac{\modu{\emm_a}^2}{8\pi \,m_{DM}^2}\delta\left(1-y\right)
\end{equation}
and therefore
\begin{equation}
	\frac{\td \Gamma_a}{\td y} = \frac{\modu{\emm_a}^2}{16\pi \,m_{DM}}\delta\left(1-y\right)\,.
\end{equation}
The integrated exclusive decay width then amounts to
\begin{equation}
	\Gamma_a = \frac{1}{16\pi \,m_{DM}}\modu{\emm_a}^2
\end{equation}
where, throughout the computations, the squared matrix elements $\modu{\emm_a}^2$ are
implicitly computed for the momenta as imposed by the energy-momentum
conservation. Hence,
\begin{equation}
	\frac{1}{\Gamma}\frac{\td \Gamma_h}{\td x}=\sum_a\int\limits_x^1\frac{1}{\Gamma_a}\frac{\td\Gamma_a}{\td y}\,D_a^h(z,\mu^2)\,\frac{\td z}{z}
	=\sum_a
	\int\limits_x^1\delta\left(1-\frac x z\right)\,D_a^h(z,\mu^2)\,\frac{\td z}{z} = \sum_a D_a^h(x,\mu^2)\equiv\frac{\td N_h}{\td x}
\end{equation}
with \(D^{p}=D^{p}_q + D^{p}_g\) being the proton fragmentation
function. The last equality holds after the singular
fragmentation functions for quarks and gluons have been properly weighted by the relative color factors. Putting all together we then find
\begin{align}
J^g_x(E,\vn)
=&
\frac{1}{4\pi}\int n_{DM}(r, \vn) \,\frac{\td \Gamma_x}{\td E}\, \td r 
=
\frac{1}{4\pi\tau}\frac{1}{\Gamma}\frac{\td \Gamma_x}{\td E}\int n_{DM}(r, \vn)\, \td r 
=\\&=\nn
\frac{1}{4\pi\tau\mdm}\frac{\td N_x}{\td E} \int \rho_{DM}(r, \vn)\, \td r.
\end{align}
Notice that we define the Galactic flux on the whole sphere as $J^g_x(E):=\int_{4\pi}\td\vn J^g_x(E,\vn)$.

\subsection{The extragalactic component} 
\label{sub:The extragalactic component}

Given our preliminary discussion at the beginning of the section, it is clear that the extragalactic contribution amounts prevalently to an isotropic neutrino flux with a subdominant proton component, at least at the highest energies we are interested in. Notice that although neutrinos are very hard to detect by current ground observatories, the high-energy neutrino flux detected in dedicated experiments can still be used to set the magnitude of the extragalactic component of the UHECR~\cite{Decerprit:2011qe,Sigl:2012tu,Alvarez-Muniz:2013mfa}.

The extragalactic proton flux can itself be divided in two components: the first is the almost isotropic low-energy proton flux as measured in running experiments, which we take to follow a broken power-law spectrum:
\begin{equation}
	J_p^{exp}(E)=\frac{1}{4\pi} J_1
	\begin{cases}
	E^{-\gamma_1} \quad 1~\text{EeV}\leq E < E_\text{ankle}\\
	E_\text{ankle}^{\gamma_2-\gamma_1} E^{-\gamma_2} \quad E_\text{ankle} \leq E < E_{GZK}\\
	E_\text{ankle}^{\gamma_2-\gamma_1} E_{GZK}^{\gamma_3-\gamma_2} E^{-\gamma_3} \quad E > E_{GZK}\\
	\end{cases}\,,
\end{equation}
where $J_1$ is the total flux normalisation at 1~EeV, $E_\text{ankle} \approx 5$~EeV is the first break (ankle) and $E_{GZK} \approx 54$~EeV is the GZK cutoff as measured by Telescope Array~\cite{Ivanov:2015,Abu-Zayyad:2013jra,AbuZayyad:2012ru}; the powers are $\gamma_1=3.3$, $\gamma_2=2.7$, and $\gamma_3=4.6$ --- similar results are obtained by the Pierre Auger~\cite{Valino:2015zdi,ThePierreAuger:2015rha} observatory. This flux decays much more rapidly than the expected flux from SHDM, and is only relevant at low energies (up to and around the GZK cutoff).

The second extragalactic proton component comes from SHDM itself. If we neglect the effect of redshift (which is a very well justified assumption) we expect the flux
\begin{equation}
	J_p^{eg}(E)=\frac{1}{4\pi}\frac{R_{GZK}}{\mdm\tau}\overline{\rho}_{DM}\frac{\td N_x}{\td E}
\end{equation}
where \(R_{GZK}\) is the physical particle horizon from which the vast majority of the flux comes from (the GZK
horizon) and $\overline{\rho}_{DM}:=\Omega_m\rho_c$ with $\Omega_m=0.23$ the DM fraction of the critical energy density $\rho_c=1.1\times10^{-5}h_0^2$~GeV/cm$^3$, and $h_0=0.67$~\cite{Ade:2015xua}.

To obtain a rough estimate of the relative weight between galactic and extragalactic contributions in the proton flux, we consider the ratio of the two fluxes on the sphere:
\begin{equation}
	\frac{J_p^{eg}(E)}{J_p^g(E)}\approx\frac{4\pi R_{GZK}\overline{\rho}_{DM}}{\int\td\Omega\td r \rho_{DM}}\approx2.5\times10^{-2}
\end{equation}
where we used \(R_{GZK} = 100\)~Mpc; for the local density $\rho_{DM}$ we took the Navarro-Frenk-White profile (see below for details).  One can also easily show that even when looking towards the Galactic anticentre, the local DM contribution is about a factor of 10 stronger than the cosmological contribution.  Similar figures hold for the photon fluxes as well\footnote{It is possible to refine the estimate by including a geometric factor that accounts for the distribution of DM in the halo. For the usual DM profiles considered in the literature this correction amounts to an $\mathcal{O}(1)$ coefficient.}. This rough estimate also makes it clear that redshift effects can safely be neglected at energies above the GZK cutoff when considering photons and proton fluxes.

Neutrinos, on the other hand, do not possess a GZK horizon, and this implies that galactic and extragalactic fluxes are comparable at all energies.  To estimate the extra-Galactic neutrino flux we use the result in~\cite{Gondolo:1991rn}
\begin{equation}
	J_\nu^{eg}(E)=\frac{1}{4\pi}\frac{\overline{\rho}_{DM}}{\mdm \tau}\int\limits_0^{z_{cut}}\,\td z\modu{\frac{\td t}{\td z}}\frac{\td N_\nu}{\td E}{\Bigg\vert}_{(1+z)E}\,e^{-S_\nu(E, z)}
\end{equation}
where
\begin{equation}
	\frac{\td t}{\td z} = - H_0(1+z)\sqrt{(1+z)^3\Omega_m+\Omega_\Lambda}\,;
\end{equation}
and the neutrino opacity can be approximated at high redshift as
\begin{equation}
	S_\nu(E, z)
	=
	\begin{cases}
	3.5\times 10^{-11}(\Omega_m h_0^2)^{-1/2}(1+z)^{5/2}(E/\text{EeV})\times10^{-3},\quad 1\ll z\lesssim z_{eq}\\
	0.81\times10^{-8}(1+z)^2(E/\text{EeV})\times10^{-3},\quad z\gtrsim z_{eq}
	\end{cases}
\end{equation}
with \(z_{eq}=3360\), the redshift at matter-radiation equality, and $\Omega_\Lambda=1-\Omega_m$.  The actual calculation of the neutrino flux is quite involved, but for our purposes it suffices to say that it amount to about a tenth of the total flux; in practice, we can write $J^{eg}_\nu(E) = \kappa J^g_p(E)$ with $\kappa\approx0.1$.

\subsection{Reconstructing the total flux}
\label{sub:Reconstructing the total flux}

The total flux of UHECR can now be computed by summing over all particle species \(x=\{p,\nu,\gamma\}\).
\begin{equation}
	J_{tot}(E, \vn) = J^{exp}_p(E) + J^{eg}_p(E) + J^{eg}_\gamma(E) + J^{eg}_\nu(E) + J^{g}_p(E, \vn)+J^{g}_\gamma(E, \vn)+J^{g}_\nu(E, \vn)\,,
\label{eq:totflux0}
\end{equation}
where we can neglect the photon's and proton's extragalactic fluxes $J^{eg}_p(E)$ ans $J^{eg}_\gamma(E)$ as per our discussion above.  In Eq.~(\ref{eq:totflux0}) all terms but $J^{exp}_p(E)$ are due to SHDM decays; $J^{exp}_p(E)$ instead is the measured flux which dominates below the GZK cutoff (in this scenario).

We keep the low-energy flux in order to properly normalise the number of events at low energy. Under the assumption that the galactic fluxes amount to the SHDM decay products we therefore have that
\begin{equation}
	\frac{J^g_x(E,\vn)}{J^g_y(E,\vn)}\equiv\frac{D^x(E)}{D^y(E)};\qquad x,y\in\{p,\nu,\gamma\}
\end{equation}
as the whole angular dependence enters through the line of sight
integral which is common to both numerator and denominator. 

Precise studies of the fragmentation function (see~\cite{Aloisio:2003xj} and references therein) reveal that at the relevant energies
\(D^x(E)\propto E^{-1.9}\) independently of the
species \(x\). Hence, we can write \(D^x(E)=:A_x (E/\hat{E})^{-1.9}\)
where \([A_x] = [E^{-1}]\), $\hat E$ is the reference energy for the normalisation of the spectrum, and rather generically~\cite{Aloisio:2003xj}:
\begin{equation}
	\frac{A_\gamma}{A_p}=2\div3,\qquad \frac{A_\nu}{A_p}=3\div4\,.
\end{equation}
Because of the behaviour of the fragmentation functions we can then write
\begin{equation}
	J_{tot}(E, \vn) = J^{exp}_p (E)+ J^{eg}_\nu(E) + \left(1+ \frac{A_\gamma}{A_p}+\frac{A_\nu}{A_p}\right)J^{g}_p(E, \vn)
\end{equation}
where
\begin{equation}
	J^{g}_p(E, \vn) = \underbrace{\frac{A_p}{4\pi \mdm\tau}}_{:=\jcap{p}}\left(\frac{E}{\hat{E}}\right)^{-1.9}\int \rho_{DM}(r, \vn)\, \td r
\end{equation}
and \(\left[\jcap{x}\right] = [E^{-2} t^{-1}]\).

We write the \emph{integral} flux of photons, assumed to originate from SHDM decay, as
\begin{align}
\Phi_\gamma^{DM}(E_\text{cut}) & := \frac{1}{4\pi} \int\limits_{E_\text{cut}}^\infty\td E J^g_\gamma(E)
= \\\nn& =\frac{A_\gamma}{A_p}\jcap{p}
\underbrace{\int\limits_{E_\text{cut}}^\infty\td E \left(\frac{E}{\hat{E}}\right)^{-\gamma}}_{=\frac{E_\text{cut}}{\gamma-1}\left(\frac{E_\text{cut}}{\hat{E}}\right)^{-\gamma}}
\underbrace{\int_{4\pi}\td\vn\td r\, \rho_{DM}(r, \vn)}_{:=\omega}
\end{align}
with \(\gamma=1.9\).  As the ratios of \(A_x\) factor are known quantities, we can use the experimental limits on the integral photon flux~\cite{Rubtsov:2015trh,Aglietta:2007yx,Abraham:2009qb,Scherini:2013tya,Bleve:2015nut} to bound the value of \(\jcap{p}\). In particular, the strongest bound on the \emph{integral} \(\gamma\)-flux $\Phi_\gamma^{DM}$~\cite{Scherini:2013tya} constrains the relative SHDM contribution to be 2\% above the energy of \(E_\text{cut}=10^{10}\)~GeV (10~EeV), see Table~\ref{tab:frac}.

\begin{table}
	\centering
\begin{tabular}[h]{c|ccccccc|cc}
	\toprule
   & \multicolumn{7}{c}{Auger} & \multicolumn{2}{|c}{TA} \tabularnewline
  \midrule
  $E_\text{cut}$ [EeV] & 1 & 2 & 3 & 5 & 10 & 20 & 40 & 10 & 32 \tabularnewline
  \hline
  $\epsilon_\gamma$ [\%] & 0.4 & 0.5 & 1.0 & 2.6 & 2.0 & 5.1 & 31 & 6.2 & 29 \tabularnewline
  \bottomrule
\end{tabular}
\caption{Upper bounds on the fraction of photons with respect to the total UHECR integral flux at a given energy from~\cite{Rubtsov:2015trh,Aglietta:2007yx,Abraham:2009qb,Scherini:2013tya,Bleve:2015nut}.}
\label{tab:frac}
\end{table}

Accounting for the experimental bound then implies
\begin{equation}
	\Phi_\gamma^{DM}(E_\text{cut}) \lesssim \epsilon_\gamma\Phi^{exp}(E_\text{cut})
	\implies 
	\jcap{p}\lesssim\frac{\gamma-1}{A_\gamma/A_p}\frac{1}{E_\text{cut}\omega}\left(\frac{E_\text{cut}}{\hat E}\right)^\gamma\epsilon_\gamma\Phi^{exp}(E_\text{cut})
\end{equation}
where \(\Phi^{exp}(E_\text{cut})\) is the experimental integral spectrum, i.e., the total number of events detected by an experiment per unit of area steradians and time with an energy of \(E_\text{cut}\) or greater.  Let us remark that this is a ``best case scenario'', that is, we directly employ the maximal SHDM flux currently allowed by observations.

All together, the flux from the direction $\vn$ is given by
\begin{equation}
	J_{tot}(E,\vn) = J^{exp}_p(E) + \kappa J^{eg}_p(E) + \left(1+ \frac{A_\gamma}{A_p}+\frac{A_\nu}{A_p}\right)J^{g}_p(E, \vn) \, ,
\label{eq:totflux}
\end{equation}
which integrated from a given threshold energy $E_\text{cut}$ gives the total integral flux $\Phi_{tot}(E_\text{cut},\vn)$.

\section{Methodology} 
\label{sec:Methodology}

\subsection{Dark Matter profiles}
\label{sub:Dark Matter profiles}

In order to explicitly compute the UHECR flux coming from SHDM decays we need to specify the SHDM density in our Galaxy.  As at the time of writing a number of different DM profiles is still compatible with the observations, we have run our analysis for the most common ones (see~\cite{Cirelli:2010xx}), and present here only two bracketing cases which give us the largest and smallest anisotropy.  These respectively are the Einasto profile~\cite{Einasto:1965czb}:
\begin{equation}
	\rho_{DM}(r_g) = \rho_* \, \exp\left\{ -\frac{2}{\alpha} \, \left[\left(\frac{r_g}{r_*}\right)^\alpha-1\right] \right\} \, ,
\label{eq:ein}
\end{equation}
with $r_*=35.24$~kpc, $\alpha=0.11$ and $\rho_*=0.021$~GeV/cm$^3$ and the Navarro-Frenk-White (NFW) profile~\cite{Navarro:1995iw}:
\begin{equation}
	\rho_{DM}(r_g) = \rho_* \, \frac{r_*}{r_g} \, \left(1+\frac{r_g^2}{r_*^2}\right)^{-2} \, ,
\label{eq:nfw}
\end{equation}
with $r_*=24.42$~kpc, and $\rho_*=0.184$~Gev/cm$^3$. Notice that here we used galacto-centric coordinates where $r_g^2 = (x-R_\odot)^2 + y^2 + z^2$ with $(x,y,z)$ the Cartesian set centred on the Earth --- in our numerical computations we also cut off the Galactic halo at 100~kpc.

\subsection{Observables}
\label{sub:Observables}

There are a number of different ways to look for an anisotropic signal in the sky.  In our case we expect a sharp peak in the flux in the area around the Galactic centre with respect to the Galactic anticentre.  Assuming we have access to the entire celestial sphere, the two most logical observables one can construct from the events collected within a given solid angle are the ratio and difference of the forward and backward (integral) fluxes from a given cone of aperture $\beta$.  The ratio $r_\text{fw/bw}$ is defined as:
\begin{equation}
	r_\text{fw/bw} := \frac{\Phi_\text{forward}(E_\text{cut},\beta)}{\Phi_\text{backward}(E_\text{cut},\beta)} \, ,
\label{eq:r_def}
\end{equation}
where $\Phi(E_\text{cut},\beta) := \int_0^\beta \td\vn \Phi(E_\text{cut},\vn)$ and ``forward'' and ``backward'' refer to the direction of the cone with respect to the Galactic centre.  The difference instead is defined as:
\begin{equation}
	d_\text{fw/bw} := \frac{\Phi_\text{forward}(E_\text{cut},\beta)-\Phi_\text{backward}(E_\text{cut},\beta)}{\Phi_\text{forward}(E_\text{cut},\beta)+\Phi_\text{backward}(E_\text{cut},\beta)} \, .
\label{eq:d_def}
\end{equation}
Like we said, these definitions assume that one single experiment (or a cross-calibrated combination of more than one~\cite{Aab:2014ila,Deligny:2015vol}) is able to collect events from the entire celestial sphere.  In reality, for current Earth-bound experiments it is typically possible to only see a limited region in the sky: variations on the previous definitions which can be used in this case are
\begin{equation}
	r_\text{fw/np} \equiv \frac{\Phi_\text{forward}(E_\text{cut},\beta)}{\Phi_\text{north}(E_\text{cut},\beta)} \, ,
\label{eq:alt_def}
\end{equation}
where the second cone is centred on the north (or south) Galactic poles, and similarly for $d_\text{fw/np}$.  Since these combinations are not as sensitive as the forward/backward options, and since we have in mind future experiments with full-sky coverage, we will focus on Eq.~(\ref{eq:r_def}) in what follows. The results obtained for~(\ref{eq:d_def}) and the alternative definitions \`a la~(\ref{eq:alt_def}) are presented in Appendix.

Another convenient way to test for the presence of a possible anisotropy in a spherical sky distribution is to decompose it in spherical harmonics.  Indeed, as any angular distribution on the unit sphere, the UHECR integral flux $\Phi(\bn)$ in a given direction $\bn$ can be decomposed in (real) spherical harmonics $Y_{\ell m}(\bn)$ as 
\begin{equation}
\label{ylm}
	\Phi(\bn)=\sum_{\ell\geq0}\sum_{ m=-\ell}^\ell a_{\ell m}Y_{\ell m}(\bn) \, .
\end{equation} 
Turning this around we obtain the harmonic coefficients
\begin{equation}
\label{alm}
	a_{\ell m}:=\int\td\bn Y_{\ell m}(\bn) \Phi(\bn) \, .
\end{equation} 
From the $a_{\ell m}$ coefficients one can define a direction-independent angular power spectrum $C_\ell$: 
\begin{equation}
\label{cell}
	C_\ell:=\frac{1}{2\ell+1}\sum_{ m=-\ell}^\ell \left|a_{\ell m}\right|^2 \, .
\end{equation}
The quantities $4\pi C_\ell$ characterise the amplitude (squared) of a relative flux deviation from an isotropic sky.  Likewise, one can focus on a similar decomposition but in a single coordinate (for example right ascension $\alpha$), defining the harmonic coefficients on a circle as 
\begin{equation}
\label{cndef}
	a_n := \frac12 \int\td\bn \Phi(\bn) Y_n(\alpha) \, , \quad \text{with} \; Y_n(\alpha) := \frac{1}{\sqrt{2\pi}} \, \left\{ \sqrt2\cos{n}\alpha, \, 1, \, \sqrt2\sin|n|\alpha \right\}
\end{equation}
for $\{ n>0, \; n=0, \; n<0\}$, respectively.  Similarly to the 3-dimensional case, the relative deviation from isotropy (squared) is characterised by
\begin{equation}
	c_n^2:=\frac12 \left[(a_n)^2+(a_{-n})^2\right] \, ,
\label{c2d}
\end{equation}
which is the direction-independent combination\footnote{In the case of a real experiment the flux we observe is in fact the convolution of the actual flux times the experiment's exposure: this needs to be deconvolved in order to extract the true spherical harmonics~\cite{Sommers:2000us,Aab:2014ila,Deligny:2015vol}. This can always be done numerically provided that the exposure is known, hence we do not discuss the matter any further and assume here a uniform exposure.}.

Notice that for the energies we are concerned with here (above the GZK cutoff), the impact of the Galactic magnetic field, as well as of the possible extragalactic field (yet unknown, but bounded from above~\cite{Pshirkov:2015tua,Zucca:2016iur}, is very limited and can be disregarded.  Indeed, a proton travelling through the Milky Way at 60~EeV would be deflected by at most a few degrees, see for instance~\cite{Oikonomou:2014zva}, and since the smallest aperture we consider is 5~deg, we can ignore this issue (the more so at higher energies).

\subsection{Procedure}
\label{sub:Procedure}

Our method is as follows. First of all, we choose the total number of events $\Nev$ on the sphere at 60~EeV.  The total number of events above a different energy is then given by the theoretical flux~(\ref{eq:totflux}).  We are typically interested in energies of 60~EeV and above, so we will work with 100, 300, and 500 events at this energy, with~300 being our ``realistic'' baseline value.  In~\cite{dOrfeuil:2014fsa} it was found that the new planned JEM-EUSO space observatory should be able to deliver $N_\text{ev}=250~(580)$ events with $E\geq80$~EeV within a few years of operation to reach a total exposure of $3\times10^5$~km$^2$sr\,yr (today's experiments hover around an order of magnitude less than this, but only cover a part of the sky) for the absolute energy normalisation of the Pierre Auger Observatory or Telescope Array.  At 100~EeV these numbers become~100 and~260, respectively.

We then compute the theoretical expectations for all of our observables.  In the case of $r_\text{fw/bw}$ and $d_\text{fw/bw}$ we consider opening angles ranging from 5~deg to 90~deg in steps of 5~deg.  We also vary the threshold energy from 60~EeV up to 1000~EeV for all observables.

After that, we obtain our errors from MC $10^6$~isotropic simulations with $\Nev$ events.  We perform as many simulations as it is needed in order for our MC errors to result from the actual number counting fluctuations rather than from artefacts due to limited number of simulations.  In the end we compute the confidence level at which an ideal experiment with $\Nev$ detected events at 60~EeV would be able to discriminate between isotropic and anisotropic fluxes by employing our observables; this confidence level is a measure of the performance of said observables.  This is our first result.

Secondly, in order to quantify the actual potential for a real experiment to reject isotropic flux over a flux generated by SHDM decay, we need to go one step further.  Since we are interested in excluding an isotropic flux, we compute the means and standard deviations of $10^5$~MC simulations with a given number of events under the hypothesis that the flux is given by SHDM decay, which gives us a mean $\bar X_\text{SHDM}$ and a standard deviation $\sigma^X_\text{SHDM}$.  Then we generate another set of $10^5$~MC sets assuming an isotropic flux, and take the median $\tilde X_\text{ISO}$ of a given observable.  Finally, by looking at the distance in terms of standard deviations from the SHDM mean, that is ${\cal N}_\sigma^{50\%} = (\tilde X_\text{ISO} - \bar X_\text{SHDM})/\sigma^X_\text{SHDM}$, we quantify the minimum significance at which a single experiment would be able to reject the hypothesis of isotropic UHECR distribution in 50\% of the cases, see~\cite{Waxman:1995dg} for more details on this approach.


\section{Results} 
\label{sec:Results}

\begin{figure}
\begin{center}
  \includegraphics[width=\textwidth]{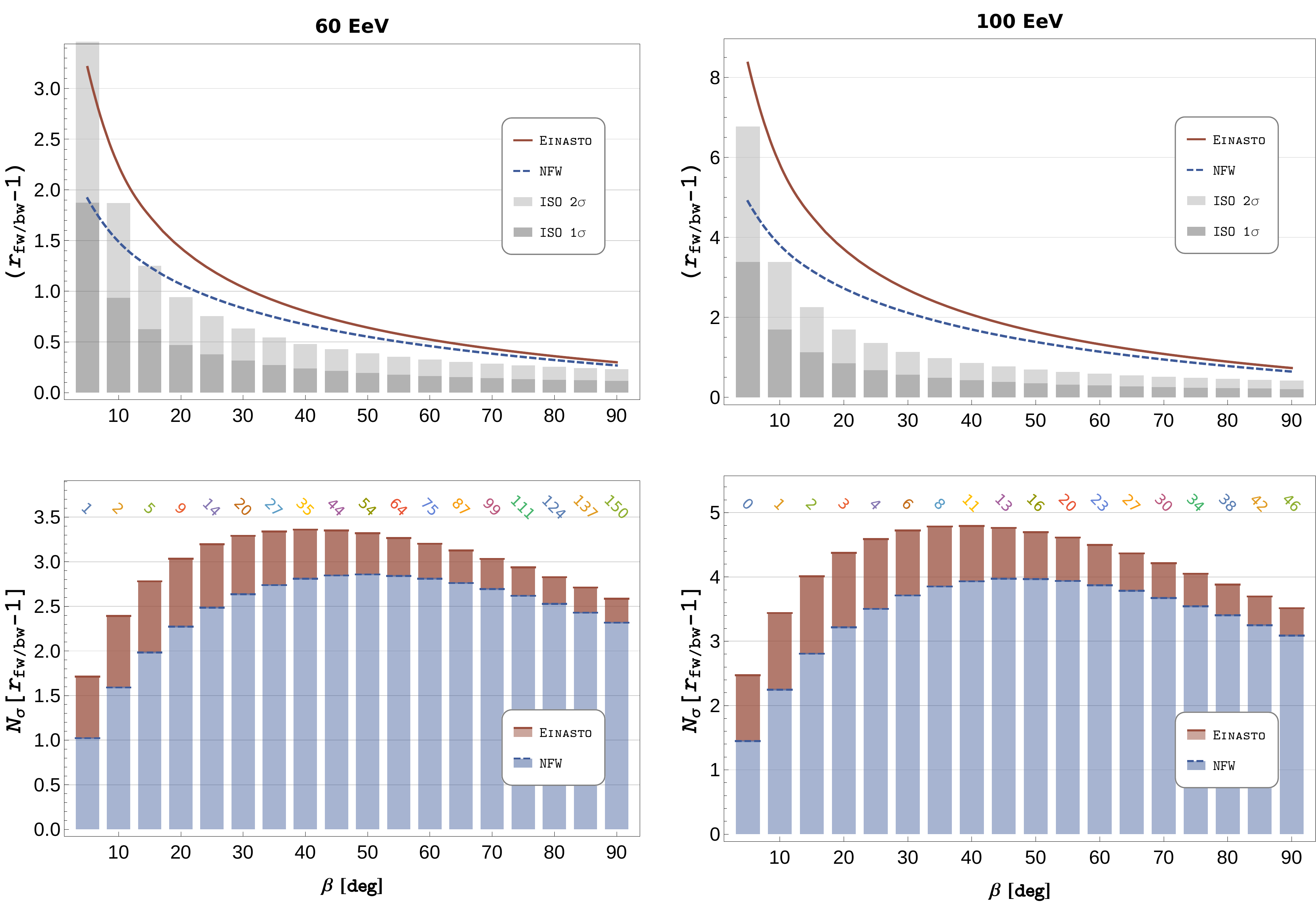}
\end{center}
\caption{Performance of $(r_\text{fw/bw}-1)$ for $\Nev=300$; see the text for details.}
\label{fig:rB300}
\end{figure}

In Figure~\ref{fig:rB300}, top panels, we present the expected theoretical $(r_\text{fw/bw}-1)$ for the two profiles NFW --- Eq.~(\ref{eq:nfw}) --- and Einasto --- Eq.~(\ref{eq:ein}) --- for varying opening angle $\beta$ in solid red and dashed blue, respectively, and the $1\sigma$ and $2\sigma$ dispersions around zero, which is the mean value expected from an isotropic distribution in dark and light grey, respectively.  We perform this analysis for threshold energies of 60~EeV (left panels) and 100~EeV (right panels).  The bottom panels show the performance of the $(r_\text{fw/bw}-1)$ observable in telling isotropy and SHDM apart again for NFW and Einasto profiles in solid red and dashed red, respectively; this perfomance is calculated as the distance between the theoretical SHDM and isotropic values in number of sigmas.  The total number of events on the sphere above 60~EeV is 300; this number reduces to 92 above 100~EeV.  For reference, the (rounded) number of events expected for an isotopic distribution within a solid angle with opening $\beta$ is also included in the bottom panel over the histograms.

The results in Figure~\ref{fig:rB300} then indicate that the opening angles most suitable for detecting possible deviations in the UHECR flux from isotropy, in the context of SHDM decay products, are intermediate values around 40~deg.  Indeed, for smaller $\beta$ the ratio between forward and backward fluxes increases dramatically, but the number of events decreases and the fluctuations become very large, especially since the overall number of events is quite small (only 92 at 100~EeV).  On the other hand, at very large $\beta$ even though we collect more statistics, the signal is diluted by a factor of about 10 with respect to the theoretical prediction for $(r_\text{fw/bw}-1)$ at 5~deg and 90~deg.  At its best, this observable can potentially tell SHDM apart from isotropy at a significance of about 3$\sigma$ for a 60~EeV cutoff, and up to about 4$\sigma$--5$\sigma$ by picking only events above 100~EeV, where $J^{exp}_p(E)$ is negligible and the SHDM flux fully dominates. For higher energies the ratio of fluxes saturates as above 100~EeV only the SHDM flux is present; however, the overall statistics rapidly degenerates and the signal is bogged down by low statistics.

\begin{figure}
\begin{center}
  \includegraphics[width=\textwidth]{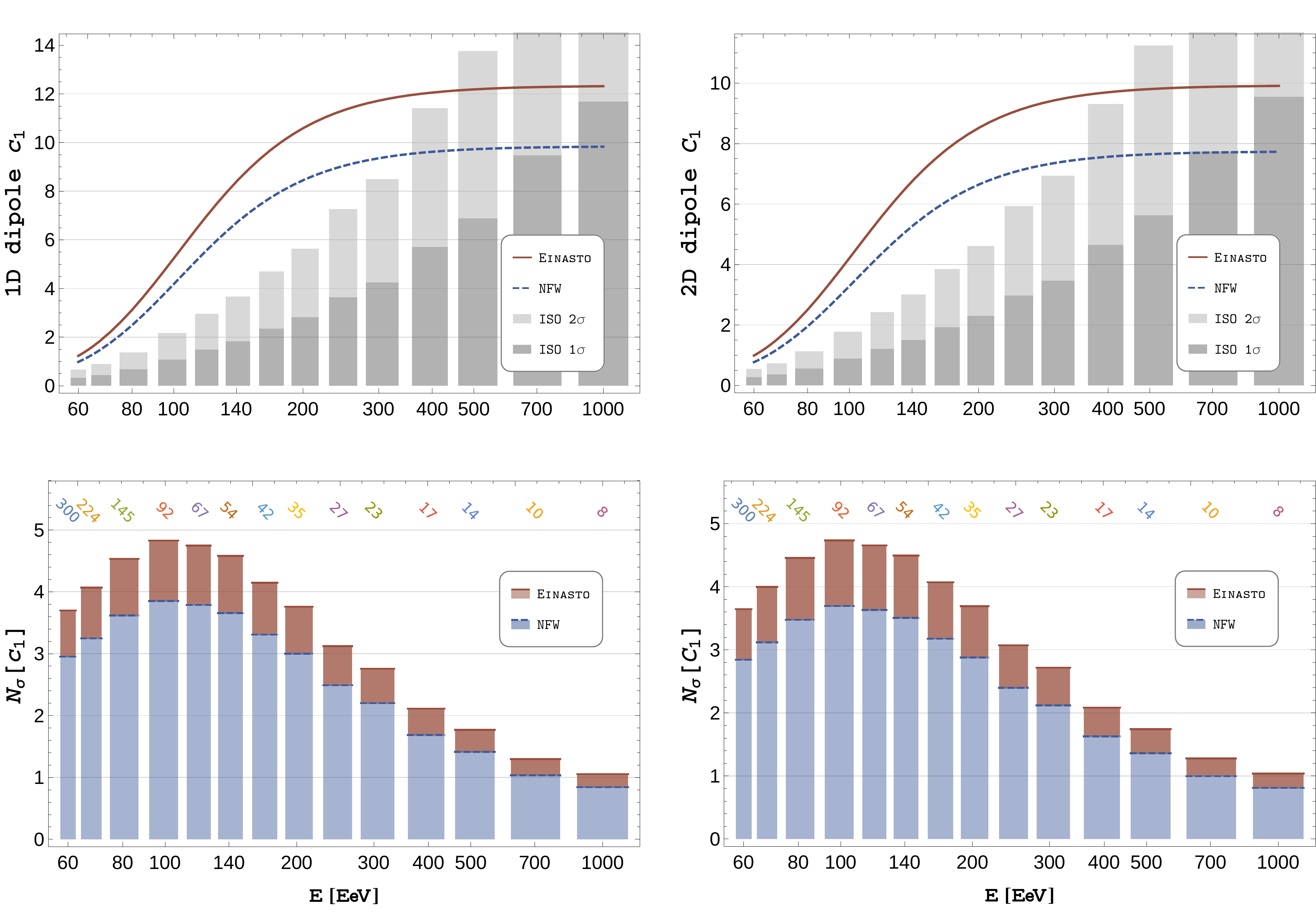}
\end{center}
\caption{Performance of the 1D dipole $c_1$ (left) and 2D one $C_1$ (right) for $\Nev=300$; see the text for details.}
\label{fig:dipo300}
\end{figure}

In Figure~\ref{fig:dipo300}, top panels, we present the expected theoretical one-dimensional, or circular (along right ascension), dipole $c_1$ (left) and two-dimensional, or spherical, dipole $C_1$ once again for the two profiles Eq.s~(\ref{eq:nfw}) and~(\ref{eq:ein}) in solid red and dashed blue, respectively, for different energy thresholds $E_\text{cut} = \{$60, 67, 80, 100, 120, 140, 170, 200, 250, 300, 400, 500, 750, 1000$\}$~EeV.  In the same plots we include the histograms of the $1\sigma$ and $2\sigma$ dispersions around zero in dark and light grey, respectively.  The bottom panels, similarly to Fig.~\ref{fig:rB300}, show the performance of the 1D and 2D dipoles to tell isotropy and SHDM apart again for NFW and Einasto profiles in solid red and dashed red, respectively.  For reference, the total (rounded) number of events is also included in the bottom panel over the histograms --- notice how the number of events drops to a total of only 23 events already at 300~EeV.

These results demonstrate that the best threshold energy for the rejection of the isotropic hypothesis against a SHDM-driven local flux is around 100~EeV, as anticipated above.  At this energy the dipoles can distinguish between these two hypotheses at a significance of about $4\sigma$ ($5\sigma$) for an NFW (Einasto) SHDM profile.  At higher energies the overall dipole magnitude reaches a plateau, but as $N_\text{ev}$ rapidly shrinks, the performance of the dipoles worsens significantly (see the top panels).

\begin{figure}
\begin{center}
  \includegraphics[width=\textwidth]{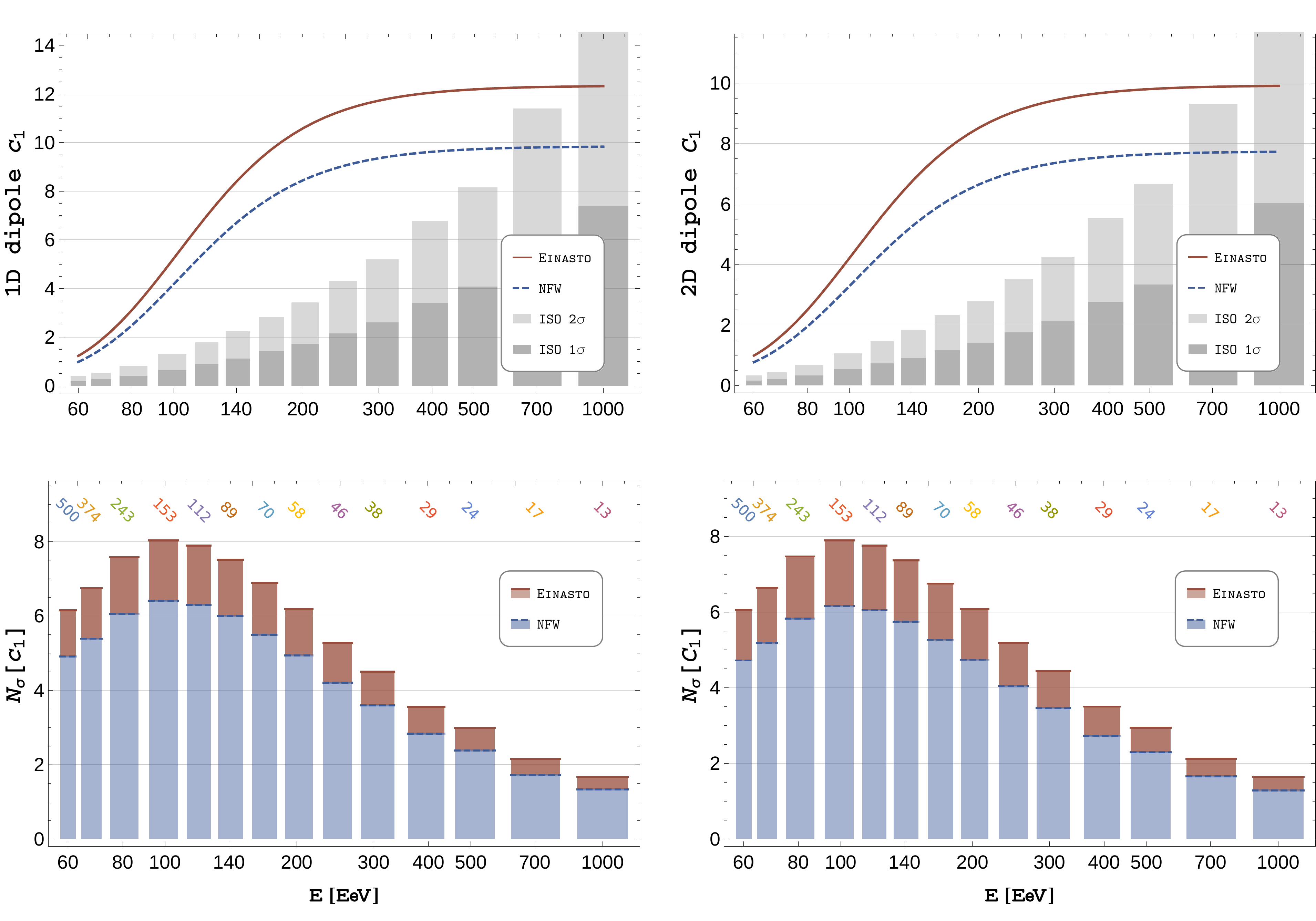}
\end{center}
\caption{Performance of the 1D dipole $c_1$ (left) and 2D one $C_1$ (right) for $\Nev=500$; see the text for details.}
\label{fig:dipo500}
\end{figure}

In Fig.~\ref{fig:dipo500} we show how the expected sensitivities change when $\Nev$ raises up to~500.  The peak sensitivities rise from around $4\sigma$ for our baseline $\Nev=300$ up to the $6\sigma$--$8\sigma$ range with as many as 500~events.  The peak sensitivities are as usual reached around 100~EeV, since, like we mentioned in the main text, from this point on the SHDM flux completely dominates, so the anisotropy does not grow anymore; on the other hand at higher energies the overall number of events reduces, boosting the counting errors.  It is interesting to see how the spherical harmonics, capitalising on the maximal number of events, become much more sensitive than ``geometric'' observables such as the ratio $(r_\text{fw/bw}-1)$ and the difference $d_\text{fw/bw}$; for these quantities a small cone/cup means small statistics, but a large one significantly dilutes the effect, for there is no weighting of the directions (as instead is the case for the spherical harmonics) --- we collect all further results for these quantities in the Appendix.

All together, we find that the dipoles and the forward-to-backward ratio are the strongest observables among the ones we have introduced in Sec.~\ref{sub:Observables} at $\Nev=300$; the harmonics however overpower the forward-to-backward ratio with larger datasets; we dig more in detail and present results for all other observables as well as the impact of $\Nev$ in the Appendix.

As promised, we also report the actual power for a real experiment to \emph{reject isotropic flux over a flux generated by SHDM decay} for the strongest observables we identified, that is, the dipoles of Fig.~\ref{fig:dipo300}.  The values of $\mathcal{N}_\sigma^{50\%}$, that quantify the minimum confidence level at which an experiment can reject the isotropic flux hypothesis, are collected in Table~\ref{tab:2}. Once again, we can see that the most favourable energies are intermediate ones, around 100~EeV to 150~EeV.

\begin{table}
	\centering
\begin{tabular}[h]{cc|cc|cc}
	\toprule
  & & \multicolumn{2}{c|}{NFW} & \multicolumn{2}{c}{Einasto} 
\tabularnewline
  $E_\text{cut}$ [EeV] & $N_\text{ev}$ & $c_1$ & $C_1$ & $c_1$ & $C_1$ \tabularnewline
  \midrule
  60 & 300 & 2.6 & 2.9 & 2.6 & 3.0 \tabularnewline
  67 & 224 & 2.7 & 3.1 & 2.8 & 3.2 \tabularnewline
  80 & 146 & 3.0 & 3.4 & 3.0 & 3.4 \tabularnewline
  100 & 92 & 3.2 & 3.8 & 3.2 & 3.7 \tabularnewline
  120 & 67 & 3.3 & 4.0 & \textbf{3.3} & \textbf{3.8} \tabularnewline
  140 & 54 & \textbf{3.4} & \textbf{4.1} & \textbf{3.3} & \textbf{3.8} \tabularnewline
  170 & 42 & 3.3 & 4.0 & 3.2 & \textbf{3.8} \tabularnewline
  200 & 35 & 3.2 & 3.9 & 3.1 & 3.7 \tabularnewline
  250 & 27 & 3.0 & 3.6 & 2.8 & 3.4 \tabularnewline
  300 & 23 & 2.8 & 3.4 & 2.7 & 3.2 \tabularnewline
  400 & 17 & 2.5 & 3.0 & 2.4 & 2.8 \tabularnewline
  500 & 14 & 2.2 & 2.7 & 2.1 & 2.5 \tabularnewline
  700 & 10 & 1.9 & 2.3 & 1.8 & 2.1 \tabularnewline
  1000 & 8 & 1.7 & 2.0 & 1.6 & 1.9 \tabularnewline
  \bottomrule
\end{tabular}
\caption{Potential ${\cal N}_\sigma^{50\%}$ for a single experiment to observe departure from isotropy via the 1D and 2D dipoles, in terms of the confidence with which such hypothesis would be rejected in half of such experiments.}
\label{tab:2}
\end{table}


\section{Conclusion} 
\label{sec:Conclusion}
In this paper we discussed the future prospects for the detection of an anisotropic UHECR flux generated by SHDM decays.  After having identified observables tailored to test the hypothesis of an isotropic UHE proton flux, we assessed their potential via MC simulations that spans the reach of near-future and next-generation experiments, in the best case scenario for which the SHDM flux is maximal.  The strongest observables are the 1D and 2D harmonic dipoles (the 1D dipole is obviously along right ascension), and the forward-to-backward ratio.  These quantities can potentially discriminate between isotropic and anisotropic fluxes at the 4$\sigma$ to 5$\sigma$ confidence level for NFW and Einasto SHDM profiles, respectively.  The best values hold for an energy cutoff at (and around) 100~EeV, and, in the case of $(r_\text{fw/bw}-1)$, for opening angles around 40~deg --- smaller angles suffer from a sharp decrease in statistics.  As the total number of events $\Nev$ grows, the spherical harmonics tend to outperform the other observables.  Finally, we would like to remark that, given a measurement of a dipole, the entire harmonic power spectrum can be predicted: higher multipoles, albeit less pronounced, can therefore be utilised to discern the SHDM origin of a potential anisotropic flux component.

Moreover, we have also quantified the power of a single experiment to reject the isotropy hypothesis by employing our best observables.  In this regard we find that, by looking at the integral flux above 100~EeV, the significance at which the isotropic flux hypotesis can be refuted exceeds the $3\sigma$ level in at least 50\% of the cases.

To conclude, UHECR experiments do possess a significant potential to test models of SHDM, the phenomenology of which is not accessible at collider and direct detection experiments.  Our results can be used straightforwardly within concrete SHDM models to derive distinguishing features that can potentially by observed in the upcoming UHECR sky.

\section*{Acknowledgements} 
\label{sec:Acknowledgements}
LM acknowledges the Estonian Research Council for supporting his work with the grant PUTJD110; FU was supported by the ERC grant PUT808 and the ERDF CoE program.  The authors thank Lucia Perrini and Antonio Racioppi for useful discussions, and Andrew Fowlie and Peter Tinyakov for reading the draft. This paper is dedicated to the memory of the Speakeasy, in Tallinn, where most of this work was conceived. 

\appendix
\section*{Appendix: all the rest} 

\begin{figure}
\begin{center}
  \includegraphics[width=\textwidth]{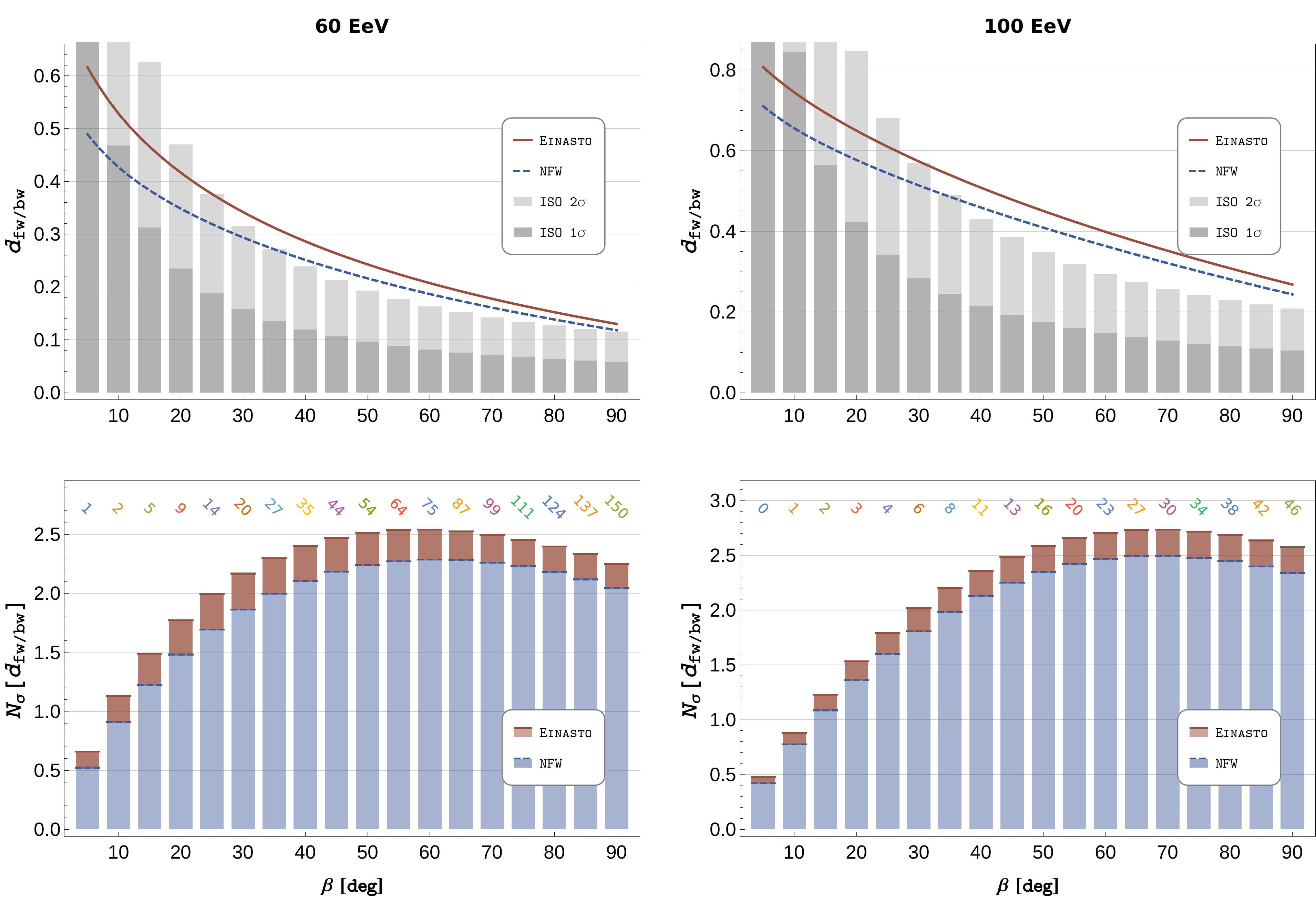}
\end{center}
\caption{Performance of $d_\text{fw/bw}$ for $\Nev=300$; see the text for details.}
\label{fig:dB300}
\end{figure}

In addition to the $(r_\text{fw/bw}-1)$ ratio of Eq.~(\ref{eq:r_def}) shown in Figure~\ref{fig:rB300}, an alternative way to look for a forward/backward flux anisotropy is to use the difference $d_\text{fw/bw}$ from Eq.~(\ref{eq:d_def}).  In Fig.~\ref{fig:dB300} we show the performance of this observable for the $\Nev=300$.  The panels are organised in the same way as in Fig.~\ref{fig:rB300} and show that this quantity is not as sensitive as the ratio $(r_\text{fw/bw}-1)$, reaching at most the $2.5\sigma$ level.

\begin{figure}
\begin{center}
  \includegraphics[width=\textwidth]{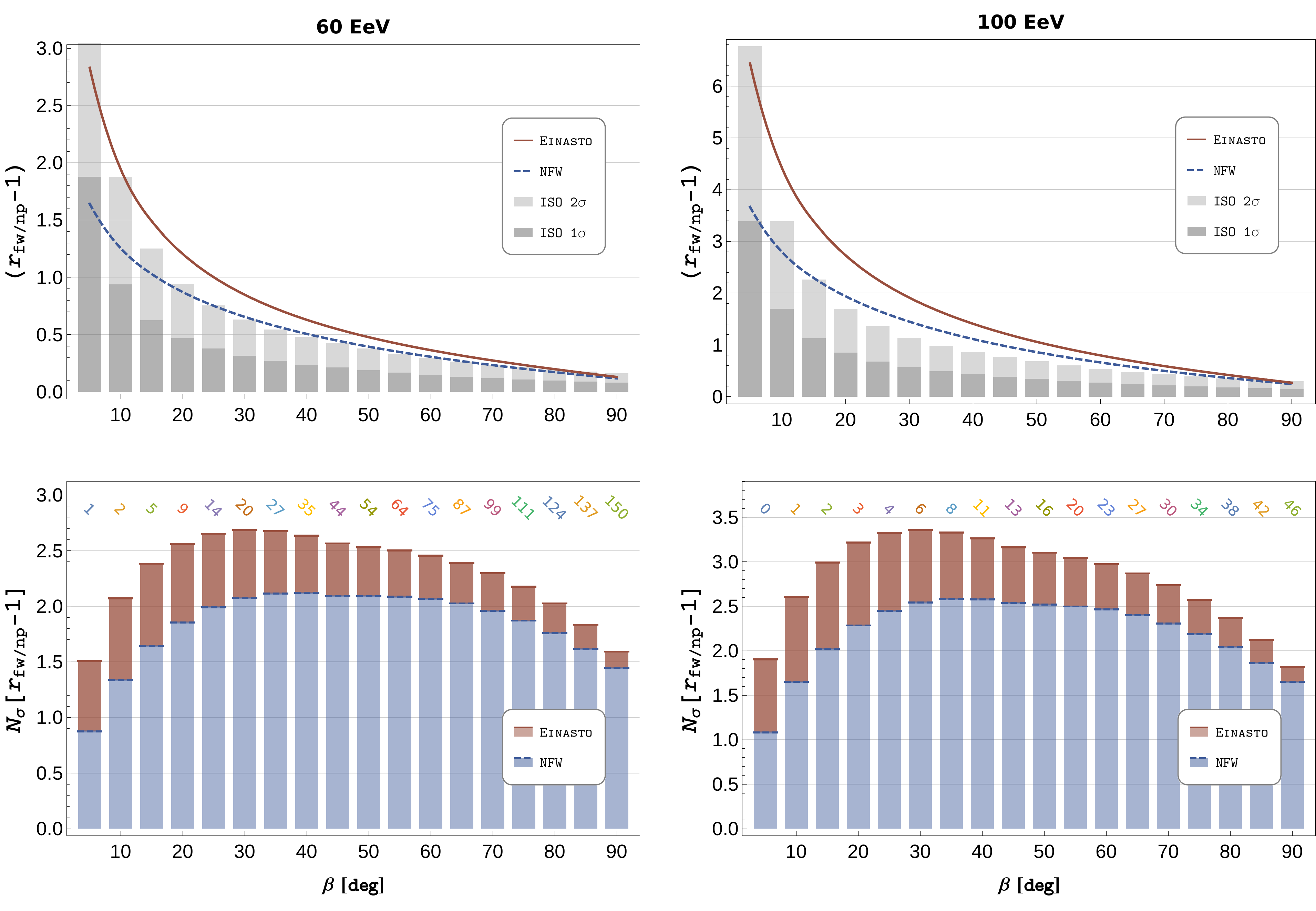}
\end{center}
\caption{Performance of $(r_\text{fw/np}-1)$ for $\Nev=300$; see the text for details.}
\label{fig:rN300}
\end{figure}

\begin{figure}
\begin{center}
  \includegraphics[width=\textwidth]{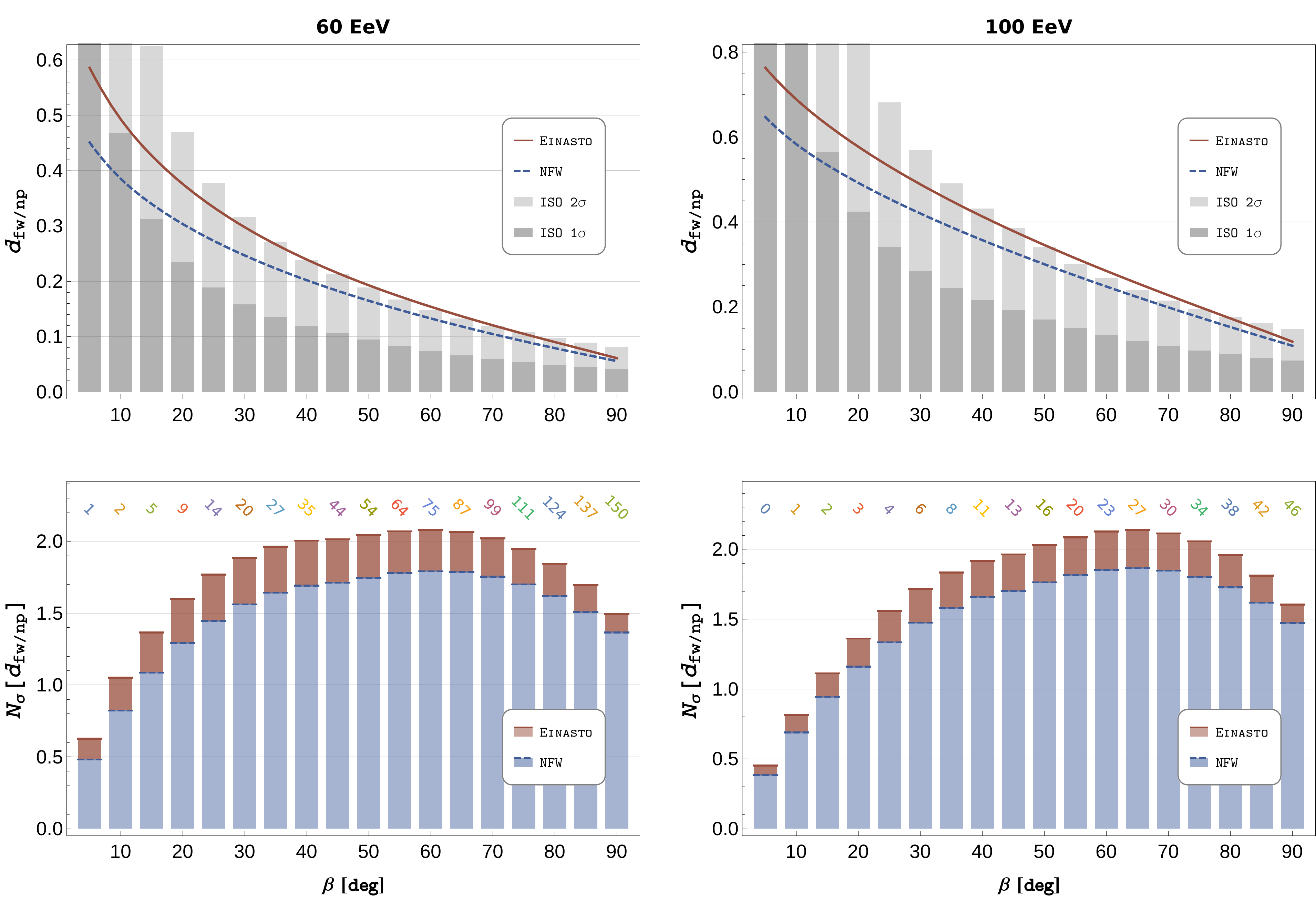}
\end{center}
\caption{Performance of $d_\text{fw/np}$ for $\Nev=300$; see the text for details.}
\label{fig:dN300}
\end{figure}

If an experiment does not have access to the forward and backward regions of the sky, one may be forced to pick instead observation cones centred towards north (or south, or even left and right which are all equivalent given the spherical symmetry of the DM density distribution).  In Figs.~\ref{fig:rN300} and~\ref{fig:dN300} we show the two ``geometric'' observables $(r_\text{fw/np}-1)$ and $d_\text{fw/np}$ using the Galactic North Pole as the ``background'' counting region.  Again, we take $\Nev=300$, and the panels are organised in the same way as in Fig.~\ref{fig:rB300}.  The $(r_\text{fw/np}-1)$ is of course less sensitive compared to the $(r_\text{fw/bw}-1)$, as is expected since a cone towards the Galactic North Pole encompasses more SHDM by volume than a corresponding cone pointing backward from the Galactic Centre.  It is also interesting to notice how the effect of the cross-correlation between forward and north cones washes out (as it should) the sensitivity to the anisotropy, as the cones overlap beyond 45~deg.  The $d_\text{fw/np}-1$ is even less sensitive, see Fig.~\ref{fig:dN300}.

\begin{figure}
\begin{center}
  \includegraphics[width=\textwidth]{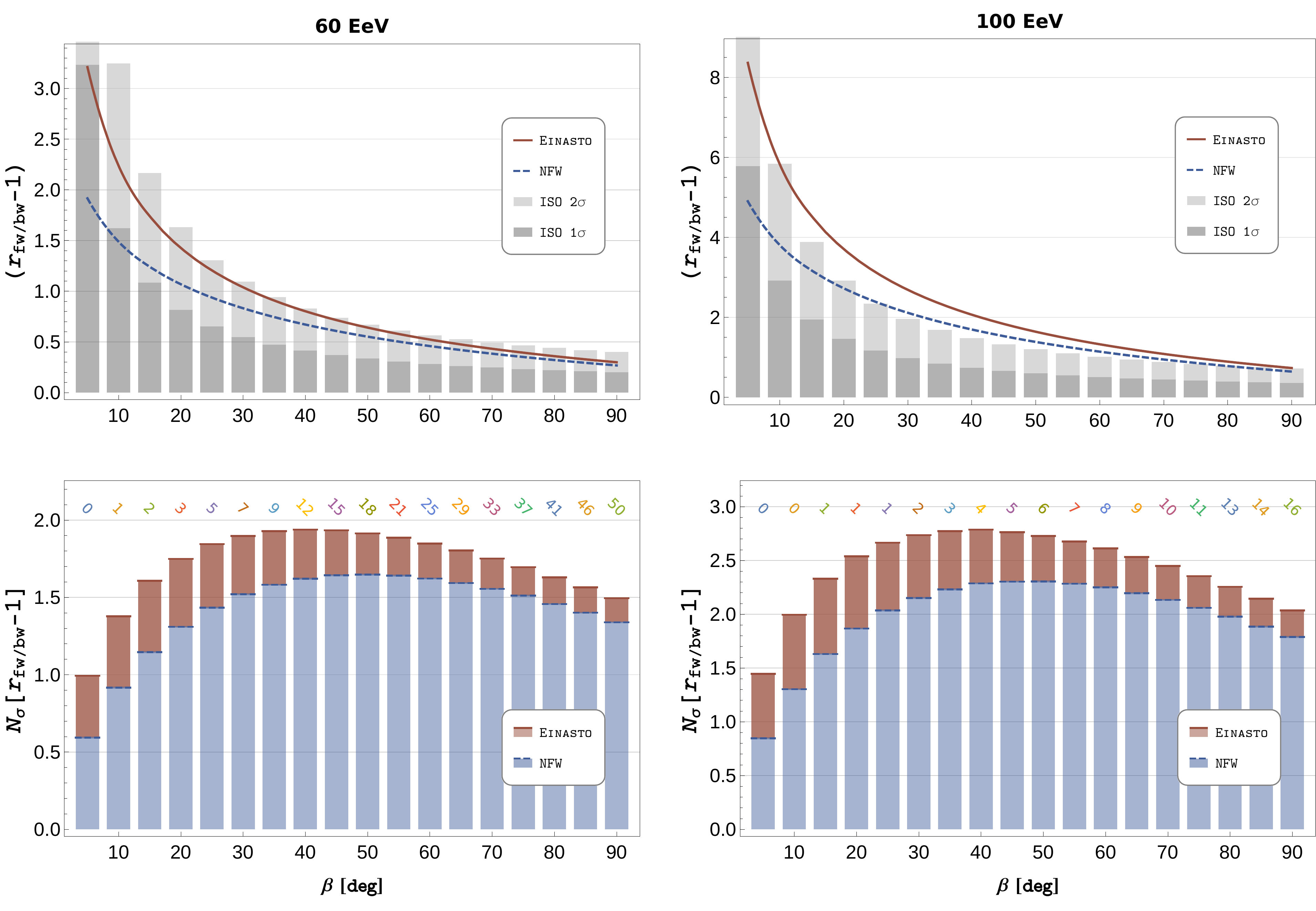}
\end{center}
\caption{Performance of $(r_\text{fw/bw}-1)$ for $\Nev=100$; see the text for details.}
\label{fig:rB100}
\end{figure}

\begin{figure}
\begin{center}
  \includegraphics[width=\textwidth]{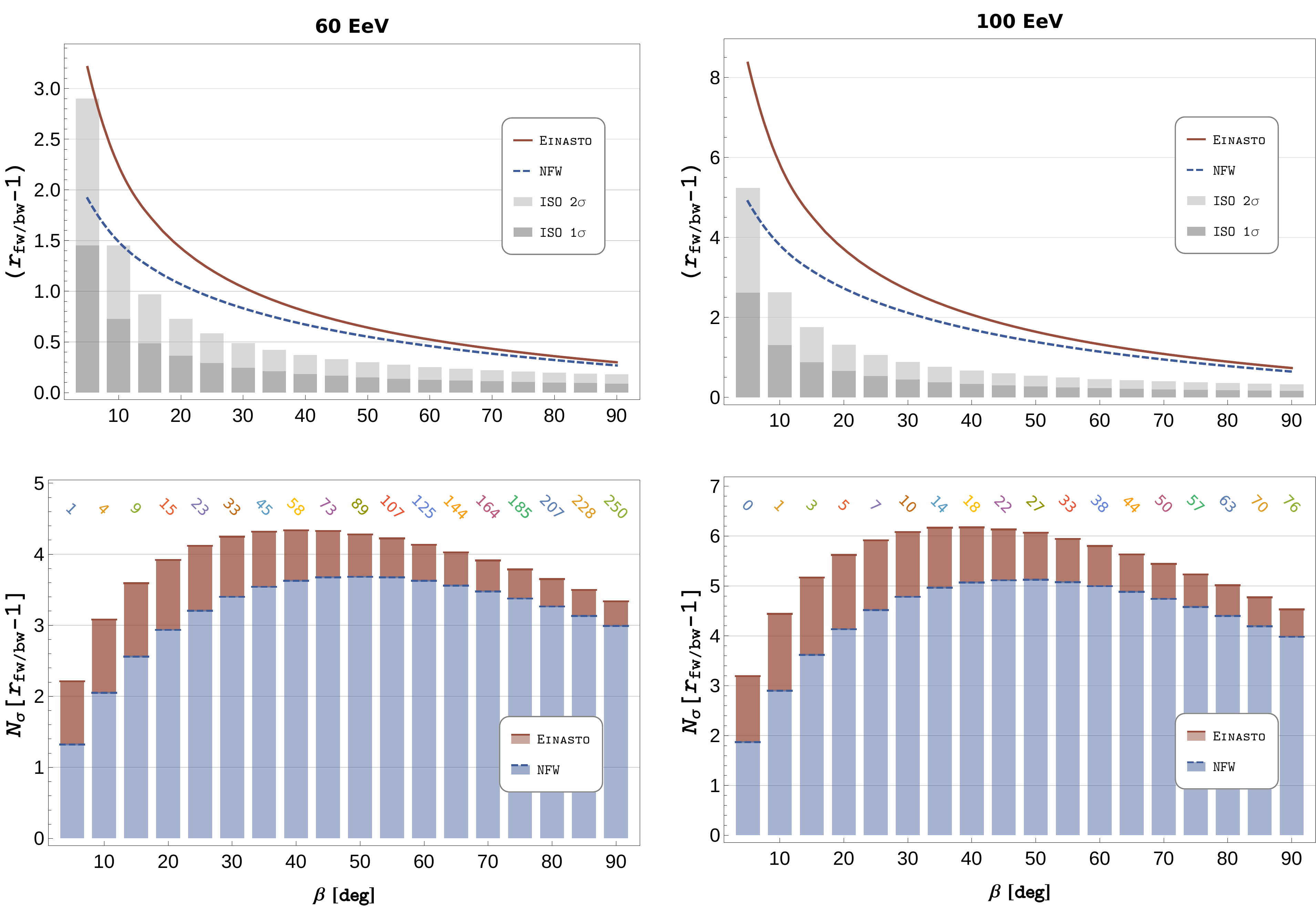}
\end{center}
\caption{Performance of $(r_\text{fw/bw}-1)$ for $\Nev=500$; see the text for details.}
\label{fig:rB500}
\end{figure}

To get a grip on the dependence of the ``geometric'' observables on the overall number of events, we have performed our simulations with $\Nev=100$ and $\Nev=500$, and the results for $(r_\text{fw/bw}-1)$ are presented in Figs.~\ref{fig:rB100} and~\ref{fig:rB500}, respectively.  With only 100~events the sensitivity is down to $1.5\sigma$--$2\sigma$ for a 60~EeV cutoff, and around $2.5\sigma$ at 100~EeV, down from around $3\sigma$ at 60~EeV and $4\sigma$--$5\sigma$ with the 100~EeV cutoff.

On the other hand, when $\Nev=500$, such as it may be expected from the new space-borne UHECRs observatory JEM-EUSO, we would be able to attain a $5\sigma$--$6\sigma$ discrimination between an isotropic distribution and a SHDM decay scenario following a NFW and an Einasto profiles, respectively, by placing a cutoff for the integral flux at 100~EeV.  The quantity $(r_\text{fw/bw}-1)$ is the most sensitive, that is, yields the best results in terms of the balance between statistics and expected signal strength, again at around 40~deg.  Notice that despite the sensitivity of the forward-to-backward ratio grows with the number of events, it does so slower than the dipoles (for the same total number of events), see Fig.~\ref{fig:dipo500}, which therefore become the quantities of choice in the quest for SHDM.

\begin{figure}
\begin{center}
  \includegraphics[width=\textwidth]{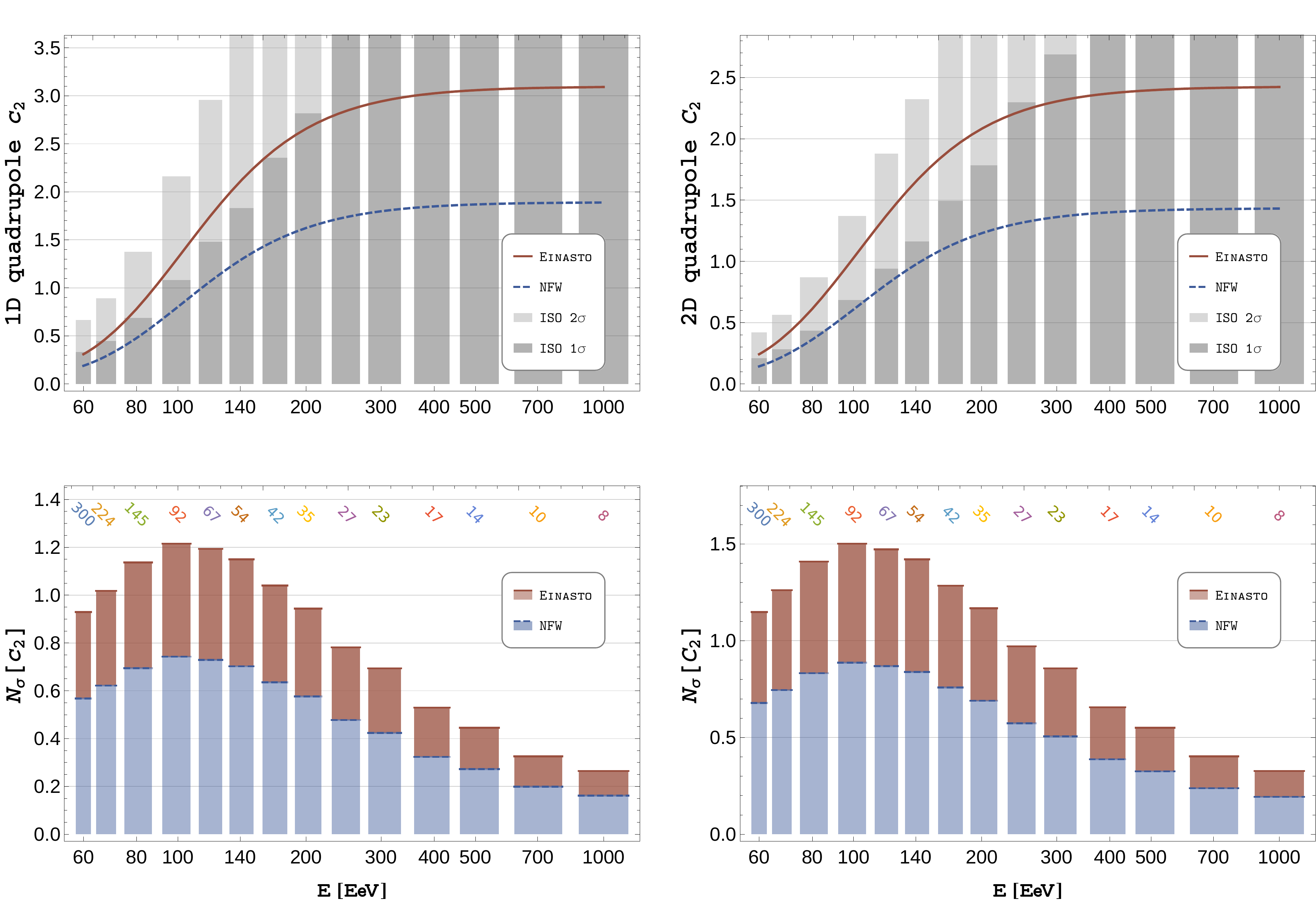}
\end{center}
\caption{Performance of the 1D quadrupole $c_2$ (left) and 2D one $C_2$ (right) for $\Nev=300$; see the text for details.}
\label{fig:quad300}
\end{figure}

Let us now consider the circular and spherical harmonics again.  Since a dipolar modulation may be explained (expected) in the UHECRs distribution by (from) scenarios other than the SHDM decay, we can look at the expected signal in higher harmonics.  In Fig.~\ref{fig:quad300} we show the one-dimensional $c_2$ (left) and two-dimensional $C_2$ (right) quadrupoles, for the reference case $\Nev=300$.  The quadrupoles are obviously not as sensitive as the dipoles, but their measurements would provide a useful cross-check on the SHDM model, because once the strength of the dipole is known we can automatically predict the strength of the quadrupole: these higher multipoles, even though less prominent, contain relevant pieces of information and can disentangle SHDM from other effects.

\begin{figure}
\begin{center}
  \includegraphics[width=\textwidth]{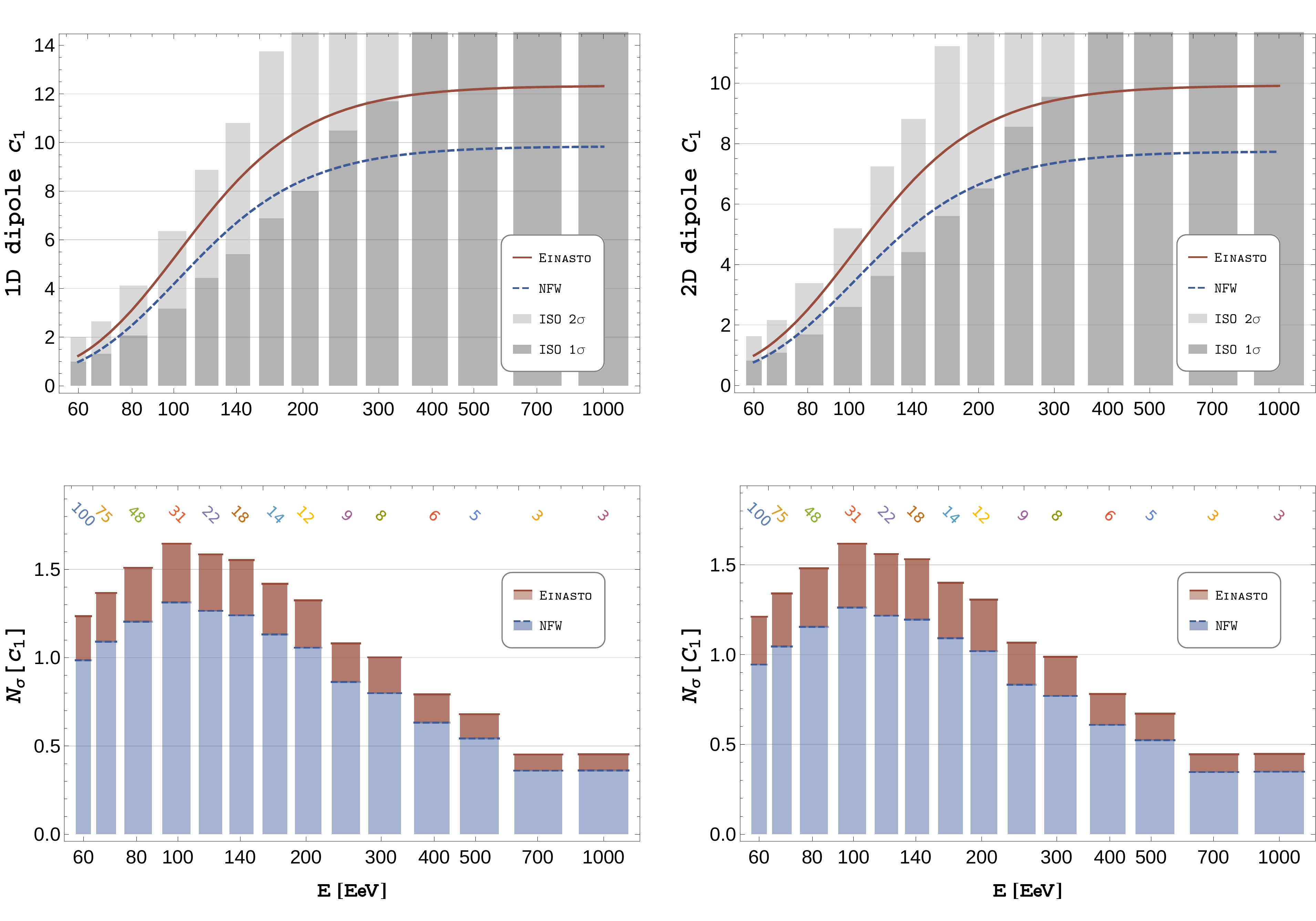}
\end{center}
\caption{Performance of the 1D dipole $c_1$ (left) and 2D one $C_1$ (right) for $\Nev=100$; see the text for details.}
\label{fig:dipo100}
\end{figure}

Finally, focussing again on the dipoles, we show in Fig.~\ref{fig:dipo100} how the expected sensitivities change when $\Nev$ drops to~100.  The peak sensitivities go from $1.5\sigma$ when $\Nev=100$, to around (and above) $4\sigma$ for our baseline $\Nev=300$ (and spike up to the $6\sigma$--$8\sigma$ with as many as 500~events, see Fig.~\ref{fig:dipo500}.  The peak sensitivities are once again reached around 100~EeV.

\bibliographystyle{hunsrt}
\bibliography{SHDM}

\end{document}